\documentclass[fleqn,10pt]{article}
\usepackage[affil-it]{authblk}
\usepackage{amsmath}
\usepackage{graphicx}
\usepackage{subfigure}
\usepackage{multibib}
\usepackage{graphicx}
\usepackage{algorithm}  
\usepackage{algorithmic}
\usepackage{float}
\usepackage{accents}
\def\degree{k}
\def\searchInfo{H}

\def\degree{k}

\title{Simplification of networks by conserving path diversity and minimisation of the search information.}
\author{H.~Yin, R.~G.~Clegg and R.~J.~Mondrag\'on}
\affil{School of Electronic Engineering and Computer Science,
Queen Mary University of London,
Mile End Road, E1 4NS, London UK}
\affil{h.yin@qmul.ac.uk}
\begin{document}
\maketitle
\begin{abstract}
Alternative paths in a network play an important role in its functionality as they can maintain the information flow under node/link failures.  In this paper we explore the navigation of a network taking into account the alternative paths and in particular how can we describe this navigation in a concise way. Our approach is to simplify the network by aggregating into groups the nodes that do not contribute to alternative paths. We refer to these groups as super-–nodes, and describe the post-aggregation network with super-–nodes as the skeleton network. We present a method to describe with the least amount of information the paths in the super--nodes and skeleton network. Applying our method to several real networks we observed that there is scaling behaviour between the information required to describe all the paths in a network and the minimal information to describe the paths of its skeleton. We show how from this scaling we can evaluate the information of the paths for large networks with less computational cost.
\end{abstract}

\flushbottom
\maketitle

\thispagestyle{empty}

\section*{Introduction}
How difficult is to navigate a city? How much information do we need to know to be able to navigate from one street to any other street? These and similar questions were studied by Rosvall~et.~al~\cite{Rosvall2005, PhysRevE.72.046117,Rosvall1118} and to answer them they introduce a new information measure, the \emph{search information}. In its simplest form, the search information relates to how many yes/no decisions a traveller has to take when navigating to reach to its destination. 
This measure has been used to study different aspect of navigability of transport networks~\cite{Barberillo2011,gallotti2016lost,Cajueiro2010,Zanin2008}, but its uses are more general, for example, recently it has been used to study task processing in the structure of topology network~\cite{Perotti2012}, brain connectome~\cite{amico2019centralized,Goni833,Avena-Koenigsberger2018,Fornito2016}, linguistic prediction~\cite{JesperLinguistic}, stability problem of wireless networks~\cite{Boushaba2017,Zayani2012}, and human behavior prediction of social networks~\cite{Shahrezaye2019}.

Our aim here is to consider a network not from the view of the traveller but from the view of the network operator. How do we describe the navigability of a network in a concise way? One of our concerns is that a network may contain many alternative paths between two nodes and we would like to capture the existence of these alternative paths. The reason to base our description on path diversity is because it plays an important role in the network robustness as alternative paths can maintain the information flow if one path is not available.

\begin{figure}
\begin{center}
\includegraphics[width= 12cm]{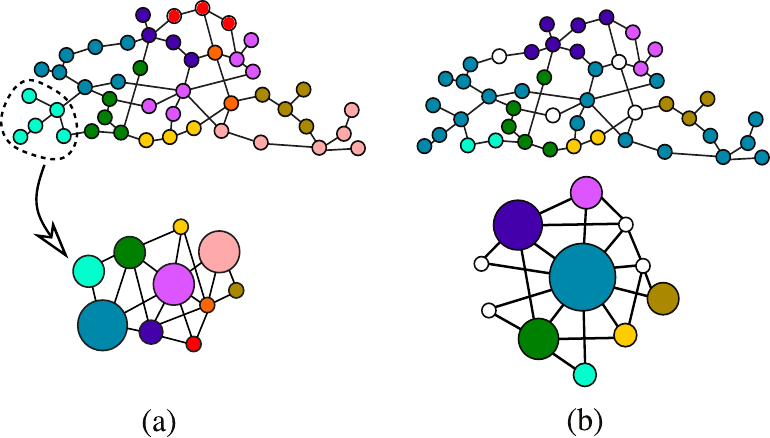}
\end{center}
\caption{\label{fig:one}
(a) Agglomeration of a set of nodes (bottom) from the original network (top).  (a)-(b) Two  skeleton networks (bottom) with different connectivities obtained from the network in (top). The size of the super--nodes is proportional to the number of nodes contain in the super--node. The skeleton networks cannot be simplified further as this would introduce multi--links.
}
\end{figure}

In part we are looking to partition the network into groups, where a group is the set of nodes where there is a unique path between the members of the group but different alternative paths between members of different groups. 
The procedure to simplify a network to a smaller network where both networks have the same number of alternative paths is based on link--contraction, that is the agglomeration of the nodes that do not contribute to alternative paths into a \emph{super--node} with the restriction that the agglomeration should not introduce multilinks (Fig.~\ref{fig:one}(a)). We called this link--contraction a \emph{tree--contraction}, as the subnetworks contained in the super--nodes are trees~\cite{Liu:2009,stanley2018compressing},
and the  network which describes the connectivity of the super--nodes, the \emph{skeleton} network~\cite{Liu:2009}. 

In general, the connectivity of the skeleton network obtained from the tree--contraction is not unique (Fig.~\ref{fig:one}(b)) as it depends on the particular order in which the contraction is carried out. 
To decide which of the possible simplified networks we should consider, we use the \emph{search information} ($\searchInfo$)~\cite{PhysRevE.72.046117,Rosvall2005,Trusina2005} which measures the information needed to route a signal between a source and destination nodes via all the shortest paths. Here we distinguish the information needed to describe the paths in a super--node ($\searchInfo_{s-node}$) from the paths in the skeleton network ($\searchInfo_{skeleton}$).  As it is easier  to navigate a network if its search information is low~\cite{Rosvall2005}, we search for the simplified network which has minimal search information, i.e. $\min (\searchInfo_{simp})$ where $\searchInfo_{simp}=\searchInfo_{skeleton}+\sum_{s-node}\searchInfo_{s-node}$.

Our approach to obtain the simplified network with minimal information is to assign random weights to the links of the network. The contraction is done by aggregating links in increasing order of their weights. Two nodes are aggregated if their aggregation does not introduces a multilink in the simplified network. The tree--contraction finishes when all the links are visited obtaining the skeleton network and super--nodes. Then the search information $\searchInfo_{simp}$ is evaluated. This process is repeated with different random seeds keeping track of the simplified network with minimal search information. The concept of search information was first introduced to consider the `hide-and-seek' problem\cite{Sneppen2005} in a network, that is how much information is needed to describe a shortest path from one node to another. It is known that a shortest path is not necessarily the path with minimal information~\cite{PhysRevE.72.046117}. Our method is not restricted by the assumption that the relevant paths are the shortest paths.

In next section, we will show how the partitions of the networks affect the minimal search information, how our method avoids the constraint of looking only at the shortest paths, and how we can approximate the search information of large networks with small computational cost.

\section*{Results}

\begin{figure}
\begin{center}
\includegraphics[width= 12cm]{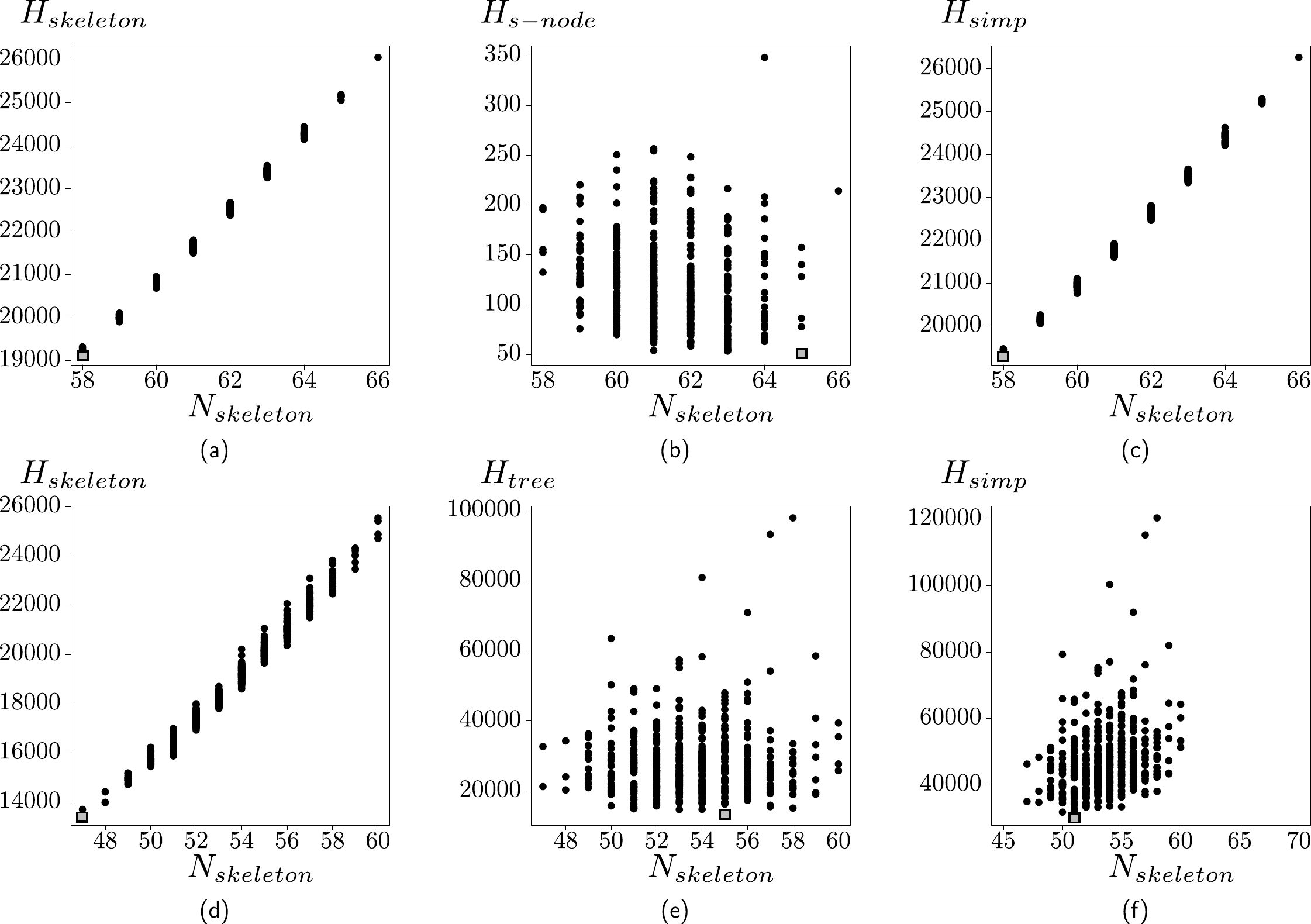}
\end{center}
\caption{\label{fig:two}
The search information of two simplified networks. (a)-(c) are for the adjacent--nouns network and (d)-(f) are for the Transport for London (TfL) network. The columns show the search information for the skeleton, super--nodes and skeleton plus super--nodes against the number of super--nodes.  Each dot in the sub--figures correspond to one of 500 simplified networks obtained by randomly selecting the contracting links. The grey squares show the minimal values.  Notice that the y--axis range in (b) is several order of magnitude smaller than in the other subfigures.}
\end{figure}

The skeleton and super--nodes both contribute to the search information of the simplified network. Fig.~\ref{fig:two} shows the search information for two real networks against the number of super--nodes. The first data set is Adjacent-Nouns network and the second data set  is the Transport for London network (TfL) describing the London underground railway network. From all the real networks that we considered (Supplementary information Tab.1), we notice that the search information of the skeleton is proportional to the number of  super--nodes (Fig~\ref{fig:two}~(a) and (d)) compared to the total search information of the super--nodes which has large variations  (Fig~\ref{fig:two}~(b) and (e)). Also, depending on the network, sometimes the main contributor to the search information comes from the skeleton network, (e.g. adjacent--nouns network,  Fig.~\ref{fig:two}(a)-(c)) and for other networks the main contribution is the information describing the super--nodes (Transport for London network in Fig.~\ref{fig:two}(d)-(f)).

\begin{figure}
\begin{center}
\includegraphics[width= 12cm]{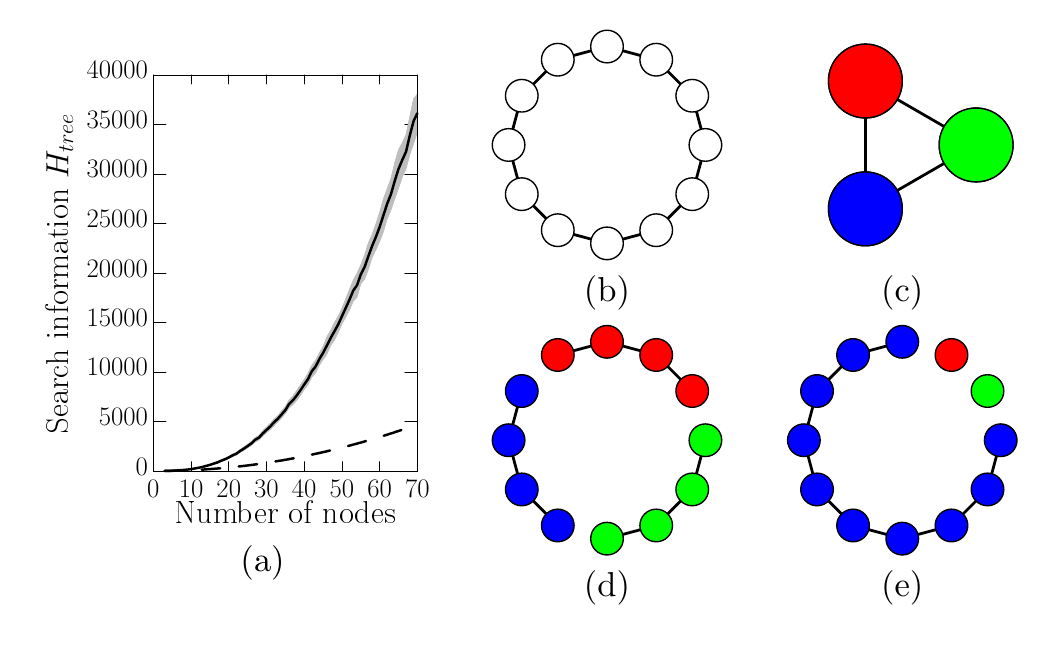}
\end{center}
\caption{\label{fig:three}  (a) Variation of the search information with the number of nodes in the tree. The black line  shows the average search information over 1000 trees and the search information is well approximated by $\searchInfo_{tree}= (0.721\pm0.019)N^{2.550\pm0.006}$ (the regression coefficient, $R^2$ is 0.999). The grey area shows one standard deviation from the average. The dashed line is the search information for the ring which grows quadratically with the number of nodes as $\searchInfo_{ring}=(N-1)(N-2)$. 
(b) A 12 node ring network will be simplified to (c) a triangular skeleton where the super--nodes of have the connectivity of a chain. (d) The minimal search information ($\searchInfo_{simp}=24$) is obtained when the nodes are distributed evenly between the super--nodes. (e) The maximal search information simplified network obtained numerically ($\searchInfo_{simp}=78$). }
\end{figure}

It is known that the search information increases with the size of the network~\cite{Barberillo2011}. The subnetwork contained inside a super--node, by construction, is a tree and we expect that the search information of these trees also increases with the number of nodes. The search information for a tree $\searchInfo_{tree}$ tends to increase as a function of the number of nodes but it would fluctuate depending on the tree connectivity. To verify the increase of $\searchInfo_{tree}$ with the number of nodes we evaluated the average search information from a random selection of connected trees with $N$ nodes. From numerical simulations (Fig.~\ref{fig:three}(a)) we observed a remarkable property, the average search information for a tree scales as $\searchInfo_{tree}\approx \alpha N^\beta$ where $\alpha=0.721\pm 0.019$ and $\beta=2.550 \pm 0.006$.

 In a network the number of nodes contained inside the super--nodes depends  on how the contraction is carried out 
 which can create large fluctuations in the number of nodes contained in the super--nodes and hence in their search information (Fig.~\ref{fig:two}(b) and (e)).
%
This large variability of the super--nodes search information can be illustrated with a ring network which is the simplest network with an alternative path  (Fig.~\ref{fig:three}(b)-(e)). In this case there are two possible routes from any node to any other node. The tree--contraction will produce a skeleton network that is a triangle. For the ring networks it is possible to show analytically (see Methods)  that the minimal search information network is when nodes of the network are evenly distributed between the three super--nodes. The other extreme, evaluated numerically, is when two super--nodes only contain one node each and the rest of the nodes are included in the third super--node, that is, larger chains have larger search information. 

\begin{figure}
\begin{center}
\includegraphics[width=12cm]{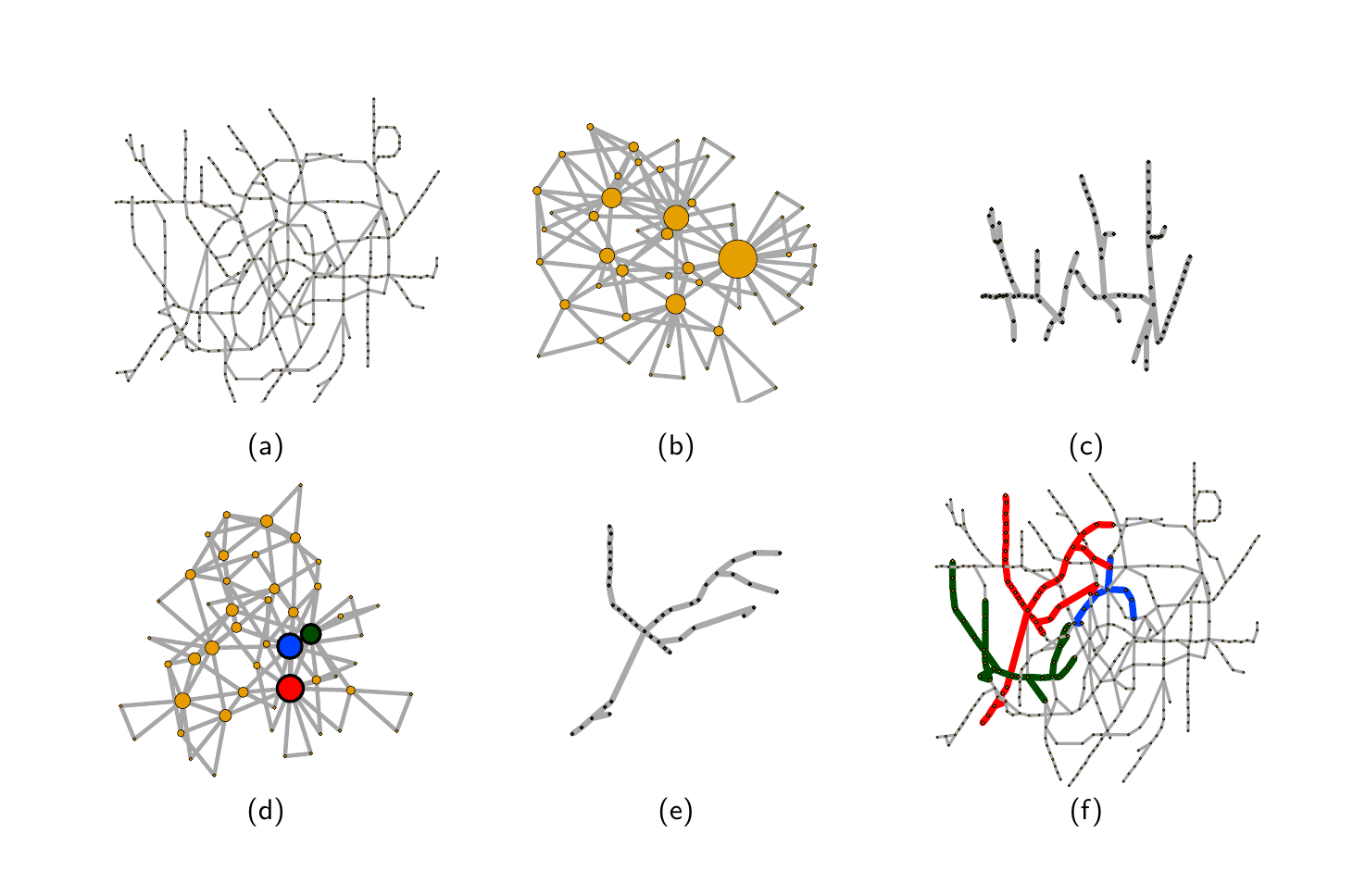}
\end{center}
\caption{\label{fig:four}
(a) The  transport for London network contracted into two skeleton networks one (b) with maximal search information and the other (d) with minimal search information. 
The largest super--nodes of these networks are very different. For the maximal search information (c) the largest super--node  contains 98 nodes and it is linked with 23 other super--nodes. For the minimal search information (e) the largest super--node contains 41 nodes and it is linked with 15 other super--nodes.  The three largest hubs marked with a thick black stroke in (d)  correspond to the set of nodes and links (thick lines)  in the original network shown in (f) .}
\end{figure}

It is known that the shortest--path is not necessarily the path with minimal search information and also it is expected that a minimal information path would tend to avoid network hubs~\cite{Rosvall2005}.  Our method  extends these observations to the general description of the network. The condition of searching for the simplified network with minimal search information produces a simplified network where super--nodes with large number of nodes tend to be avoided and the hubs of the skeleton network are now the well connected super--nodes as they are important to the path diversity. 
As an example, the TfL network (Fig.~\ref{fig:four}(a)) when simplified using the condition of maximal search information produces an skeleton network with 23 super--nodes (Fig.~\ref{fig:four}(b)) and the largest super--node contains 98 nodes  (Fig.~\ref{fig:four}(c)) compared with the minimal search information which produces a smaller skeleton of 15 super--nodes (Fig.~\ref{fig:four}(d))  and the largest super--node contains 41 nodes  (Fig.~\ref{fig:four}(e)).
The minimal search information is used to split the network into groups (super--nodes), where there is only one path between any members of a group and different paths for members of different groups. In Fig.~\ref{fig:four}(d) the red, green and blue nodes are the three largest super--nodes in the skeleton network. Fig.~\ref{fig:four}(f) shows the original network with these three super--nodes expanded to their original red, green and blue tree subgraphs.

In previous research, the search information is calculated assuming that a `traveller' follows one of the possible shortest path from the start of the traveller's walk to its destination. In here we are interested in the existence of alternative paths which not necessarily are the shortest, for example  in Fig.~\ref{fig:four}(f), a traveller has different options if she wants to go from any red station to any blue station, she can take a red-green-blue line or red-blue line route.

\begin{figure}
\begin{center}
\includegraphics[width=12cm]{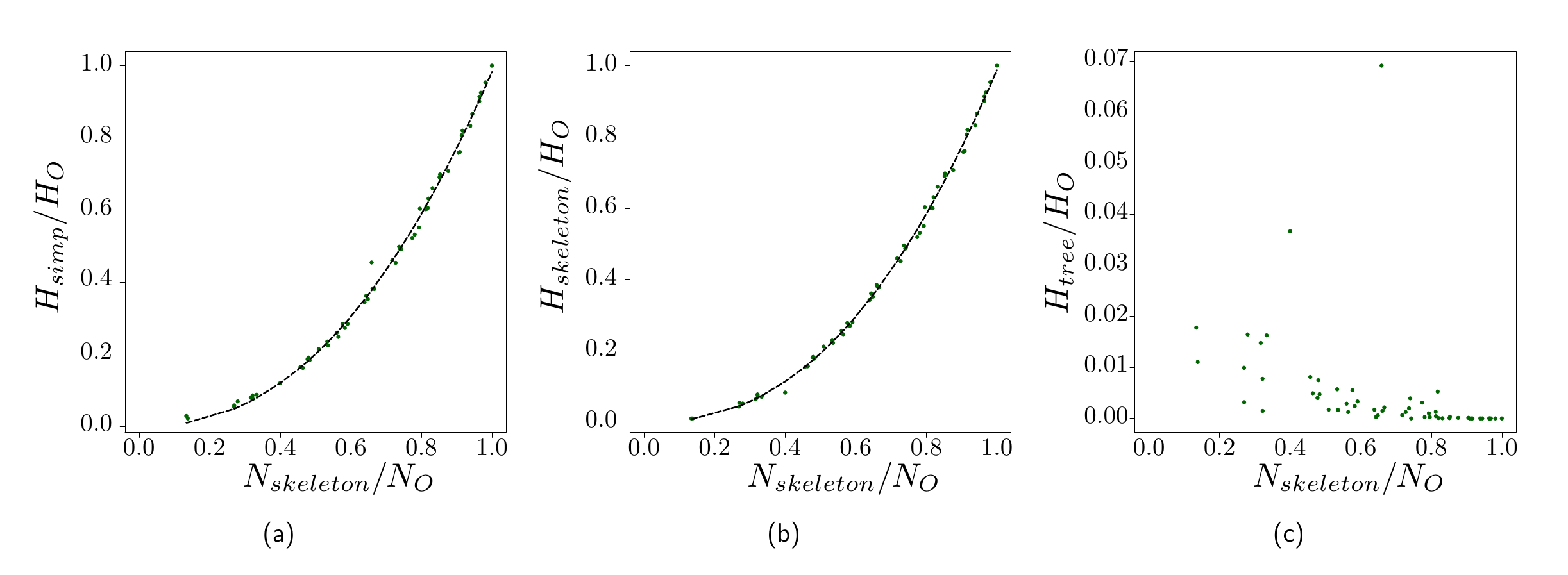}
\end{center}
\caption{\label{fig:five}
 (a) Normalised the search information of the simplified network against the number of nodes in the skeleton network for several real networks. The data is well adjusted with the black dashed curve $\searchInfo_{simp}/\searchInfo_o = (0.983\pm0.059)(N_{skeleton}/N_o)^{2.297\pm0.027}$ (the regression coefficient, $R^2$ is 0.997), where $N_{skeleton}$  and $N_o$ are the size of the skeleton network and the original network. $\searchInfo_o$ represents the search information of the original network. The normalised search information $\searchInfo_{simp}$ was separated into (b) $\searchInfo_{skeleton}$ the search information of the skeleton network  and (c) $\searchInfo_{tree}$ the search information of all the super--nodes against the number of nodes in the skeleton network. The data in (b) is well adjusted with the $\searchInfo_{skeleton}/\searchInfo_o = (0.988\pm0.004)(N_{skeleton}/No)^{2.355\pm0.021}$ (the regression coefficient, $R^2$ is 0.998).
}
\end{figure}

The minimal search information of the simplified network depends on the structure of the network. For a fully connected network the tree--contraction would not simplify the network and the search information for the original and simplified network are the same. If the original network has large chains of nodes in its structure, as these subgraphs are aggregated via the tree--contraction, the simplified network would have a small search information.  Figure~\ref{fig:five}(a) compares the ratio between the minimal search information against the search information of the network ($\searchInfo_{simp}/\searchInfo_o$) and the normalised number of nodes ($N_{skeleton}/N_o$) for many real networks. Networks like the Bison network tend to be almost fully connected and the simplified network and original network have very similar minimal search information. The other extreme is the Transport for London (TfL) network, which contains long chains in its structure. Again, as in the case of the search information for the simplified networks,  we observe a scaling behaviour for the normalised search information of the simplified network (Fig.~\ref{fig:five}(a)) and the skeleton network (Fig.~\ref{fig:five}(b)). However, there is no obvious scaling for the trees (Fig.~\ref{fig:five}(c)). The normalised search information of the original networks scales as $\searchInfo_{simp}/\searchInfo_o = (0.983\pm0.059)(N_{skeleton}/N_o)^{2.97\pm0.027}$ relative to the simplified network (Fig.~\ref{fig:five}(a)) and as $\searchInfo_{skeleton}/\searchInfo_o = (0.988\pm0.004)(N_{skeleton}/N_o)^{2.355\pm0.021}$ relative to the skeleton of the simplified networks (Fig.~\ref{fig:five}(b)). This scaling law allow us to evaluate the search information of a large network via its skeleton network. 

The evaluation of the search information can be computationally slow due to the evaluation of all the shortest paths (Dijkstra's algorithm). For large networks this process becomes slow and even slower if we need to search for a simplified network with the minimal search information. Our previous results provides a method to estimate the search information via the scaling found previously.  For example, it is 50 times faster to obtain an approximation to the search information of the Rome--road network which has over 3353 nodes and 4831 links from its skeleton network than evaluate it directly.  However in this approximation the skeleton was obtained by selecting at random the links in the tree--contraction and we cannot guarantee if the structure of the simplified network is similar to the structure of the simplified network with minimal search information. To overcome this shortcoming the contraction--tree process was modified as follows. 

\begin{figure}
\begin{center}
\includegraphics[width=12cm]{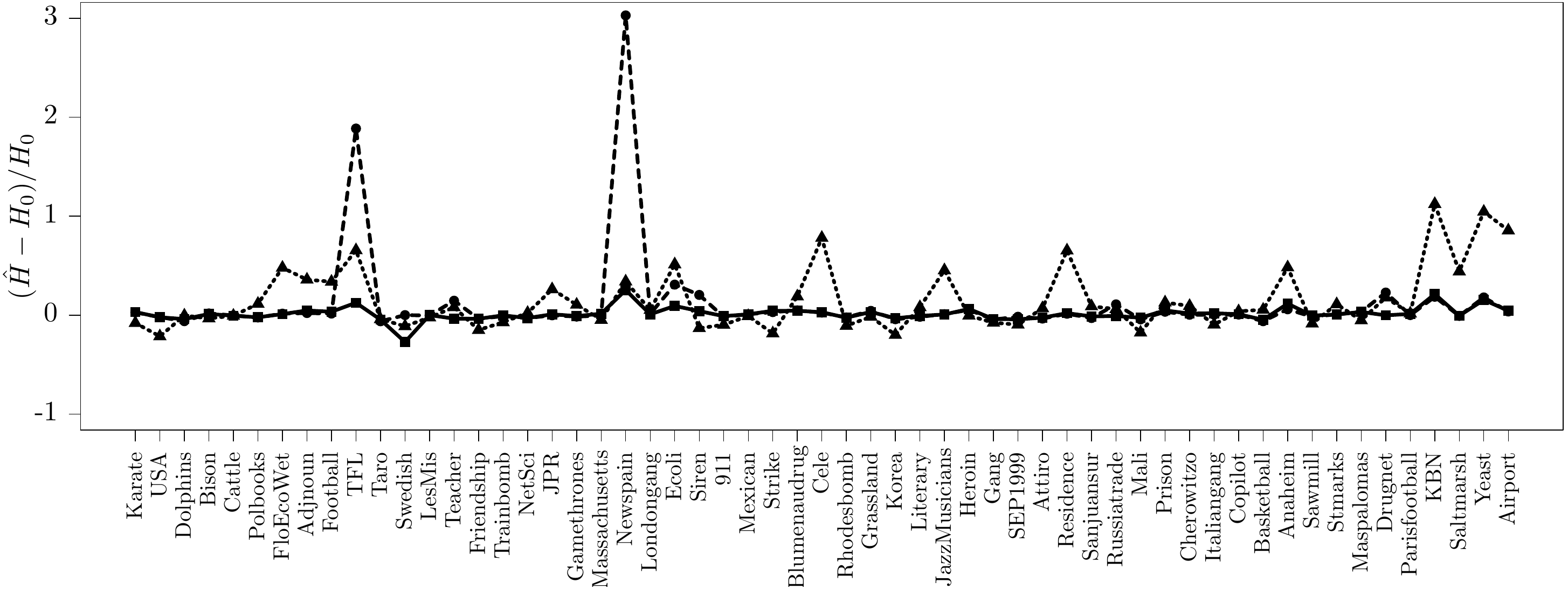}
\end{center}
\caption{\label{fig:six}
Relative error $(\hat\searchInfo-\searchInfo_o)/\searchInfo_o$ of the search information $\searchInfo_o$ when is approximated from one of the scalings $\hat\searchInfo$ described  in the main text. The best approximation is obtained using only the skeleton network of the simplified network (solid line) followed by the approximation when the skeleton is considered a tree (dotted line) and the worst approximation is for the combined skeleton and super--nodes (dashed line).
}
\end{figure}

To each link $l_{ab}$ connecting node $a$ and $b$, we assign the weight $W_{l_{ab}}=\degree_a+\degree_b$, where $\degree_a$ and $\degree_b$ are the degree of the nodes. The tree--contraction is done by contracting the links in increasing order of their weight. This strategy reduces the search information of the simplified network as it tends to aggregate chains  first. Next we consider three possible ways to approximate the search information of the original network. From the simplified network, consider the search information obtained from the skeleton network and the super--nodes, consider only the search information of the skeleton network and finally consider the search information obtained from an ``average'' tree that has the same number of nodes as the skeleton network. Figure~\ref{fig:six} shows the relative error when approximating the minimal search information of a network via the simplified network. The best approximation is obtained when using the search information of the skeleton network. 

\section*{Discussion}
The structure of a network can be studied by partitioning it into communities. Loosely speaking a community is a set of nodes which have higher connectivity to nodes within their community than nodes outside this set. It is expected that these communities reflect properties of the network, e.g. friendships in social networks. Since in this paper we are interested in the existence of alternative paths between different parts of the network,  we used a different approach to partitioning a network. Our approach is to aggregate the nodes that do not contribute to alternative paths into a group (super--node) reducing the network to a network of super--nodes (skeleton network). To decide which nodes should belong to a group we used the search information to find the paths between nodes which are described with minimal information. 

We envisage that the description of a network using our method can have applications when describing alternative paths in a communication network. The network structure inside a super--node is of a tree and the routing decision inside a tree is unique and not difficult to compute, there is only one route between two nodes in the super--node. The path diversity is captured via the skeleton network where routing decisions are made. This path diversity can be used to design maps of networks that present information in a simpler and more usable way~\cite{stanley2018compressing,lloyd2018metro}.

By searching for a simplified network via the minimal search information we obtained a partition where there is a balance between the information describing the super--nodes and the information describing the skeleton network. Remarkably, from all the networks studied here, it seems that there is a scaling of the  search information  relating the original network and the minimal search information of the skeleton of the simplified network. Even more, it seems that for some networks, this scaling can be obtained by approximating the search information of the skeleton network via the search information of an ``average'' tree. 

For large networks the simplification of a network via the minimal search information becomes computationally expensive due to the evaluation of all the shortest--paths for all pair of nodes. The scaling we observed here allows us to approximate the minimal search information for large networks from the smaller skeleton network, where, the skeleton network is obtained by doing only one tree--contraction. This tree--contraction is biased, contracting first the links where the degree of its end nodes is relatively small. This allow us to evaluate the search information of large networks with a small computational effort.

The work presented here can be extended by instead of considering the contraction of the links based on a random decision or in the degree of the nodes at the end of the link, the contraction can be based in other relevant property, for example distance or travelling time in a transport network. 

\section*{Methods}

\subsection*{The search information of networks}
Rosvall~et~al.~\cite{PhysRevE.72.046117,Rosvall2005,Trusina2005} introduced the \emph{Search Information} $\searchInfo$ to judge whether a network is difficult to navigate. This information measures the amount of information needed to route a signal from a source node to a destination node via the shortest paths. This assumes that traffic flow on a network is closely related to the shortest path.\cite{PhysRevE.72.046117}. Search Information is employed in various areas such as social networks, biological networks, computer networks etc to quantify network complexity. Let $\ell(s,d)$ be a set of linked nodes describing the shortest path from source $s$ and ending at destination $d$. 
The probability that this path is followed by a random walker who avoids exactly reversing their path is given by
\begin{equation} 
\label{eq:randomWalk}
P(\ell(s,d))=\frac{1}{k_s} \prod_{j\in \ell(s,d)/s,d} \frac{1}{k_j-1},
\end{equation}
where $j$ denotes the nodes in the shortest path $\ell(s,d)$ excluding the source $s$ and destination $d$ nodes and $k_j$ is the degree of the node $j$.  
In Eq.~(\ref{eq:randomWalk}), the
%
probability of choosing the correct link at the starting node $s$ with degree $k_s$ has probability $1/k_i$ (as there are $k_s$ possible links to choose from). For any other node in the shortest path, with the exception of the destination node, the probability of choosing the correct link when in node $j$ is $p_{j}=1/(k_j-1)$ as at it is assumed that the random walker does not retrace to the last node visited. As there can be many shortest paths between the source and destination pair, the probability to locate node $d$ using a shortest path is $P(s\rightarrow d)=\sum_{\{\ell(s,d)\}}P(\ell(s,d))$, where the sum is over all possible shortest paths $\ell(s,d)$ from $s$ to $d$. The search information from $s$ to $d$ is defined as~\cite{PhysRevE.72.046117,Rosvall2005,Trusina2005}
\begin{equation}
\searchInfo(s\rightarrow d) = -\log_2 \left(P(s\rightarrow d) \right) =-\log_2 \left(\sum_{\{\ell(s,d)\}}\frac{1}{k_s} \prod_{j\in \ell(s,d)/s,d} \frac{1}{k_j-1} \right).
\end{equation}
This information would be small if the path contains nodes of low degree or if there are many shortest paths between the source and destination nodes. The search information of the network 
is $\searchInfo_{network}=\sum_{s}\sum_{d\ne s}\searchInfo(s\rightarrow d)$, where the sum is over all source destination pairs.

\subsection*{Search information for the simplified ring network}
We consider that a simplified network consists of the skeleton network and its super--nodes.
For a ring network the tree-contraction will always produce a simplified network where the skeleton network is a triangle which connects three super--nodes (Fig.\ref{fig:three}(b)-(c)). The connectivity of the nodes forming a super--node is a chain  or a single node.
The search information of the simplified  network is $\searchInfo_{simp}=\searchInfo_{skeleton}+\searchInfo_{chain1}+\searchInfo_{chain2}+\searchInfo_{chain3}$.
The search information of the skeleton depends only on the source node which has degree 2, so $H_{skeleton} = -6\log_2(1/2)=6$, where the factor 6 is because each node can reach two of its neighbours and there are three nodes. 
The search information for a chain of $n$ nodes is 
\begin{equation}
\label{eq:chain}
\searchInfo_{chain} = (n-2)(n-1),
\end{equation}
where we used that the chain information of the two end nodes is zero and the search information for the other $n-2$ nodes is  $n-1$.
 The total search information for the ring network is $\searchInfo_{ring} = 6+(a-2)(a-1) + (b-2)(b-1)+(c-2)(c-1)$, where the value of $6$ is the search information of the skeleton network, the other terms are the search information of the three chains (Eq.~(\ref{eq:chain})), where $a$, $b$, and $c$ are the number of nodes contained in the three different chains. If $N$ is the total number of nodes in the network then $N=a+b+c$. To find the simplified network with minimal information we write $\searchInfo_{ring}(a,b)=6+(a-2)(a-1)+(b-2)(b-1)+(N-a-b-2)(N-a-b-1)$ and use the condition 
$(\partial/\partial a + \partial/\partial b)\searchInfo_{ring}(a,b)=0$.
This last expression defines the minimal search information as function of $a$ and $b$ which defines the surface $6a+6b-4N=0$ or $a=2N/3-b$. Using this value of $a$ in $N=a+b+c$ gives $c = N/3$ and using the value of $a$ in $\searchInfo_{ring}$ gives the search information as a function of only $b$ which we expressed as $\searchInfo_{ring}(b)$.  Finally the overall minimal information is defined by the derivative $\searchInfo'_{ring}(b)=0$ which gives $b=N/3$ and $a=N/3$, that is the minimal search information for the simplified ring network is when the super--nodes contain $N/3$ nodes. If $N$ is divisible by 3 then the minimal search information is $\searchInfo_{ring}=N^2/3-3N+12$. If $N$ is not divisible by 3 then nodes are divided as even as possible between the three super--nodes.

\subsection*{Path diversity of tree-contraction}
Large networks can be difficult to understand therefore there are different techniques to simplify them  leaving behind only the relevant structure. An example is the partition of a network into several clusters, in the clusters the nodes have many  connections while within cluster there are few connections. 
There are many other methods to decompose a network into clusters however they do not conserve the cyclomatic-number ($C=L-N+P$ where $N$ is the number of nodes, $L$ the number of links and $P$ the number of connected components). The tree--contraction conserves the cyclomatic number, that is the first Betti number of the graph. For comparison we used the
 Louvian method~\cite{1742-5468-2008-10-P10008} (Louvian), Fast Greedy method~\cite{PhysRevE.70.066111} (FG), Information Map method~\cite{Rosvall1118} (IM), Walk Trap method~\cite{Pons2005} (WT) and Betweenness Centrality method~\cite{Newman2004a} (BC) and evaluate the cyclomatic number of the network of clusters, which would be the equivalent to our skeleton network. 
The tree-contraction method maintains all the path diversity of the original network however other clustering methods prune edges and as a consequence the cyclomatic number of the original network and the network of clusters is different. For an example see table 2 in the supplementary information.

\subsection*{Examples of using the tree contraction to estimate the search information}
\begin{figure}
\begin{center}
\includegraphics[width=6cm]{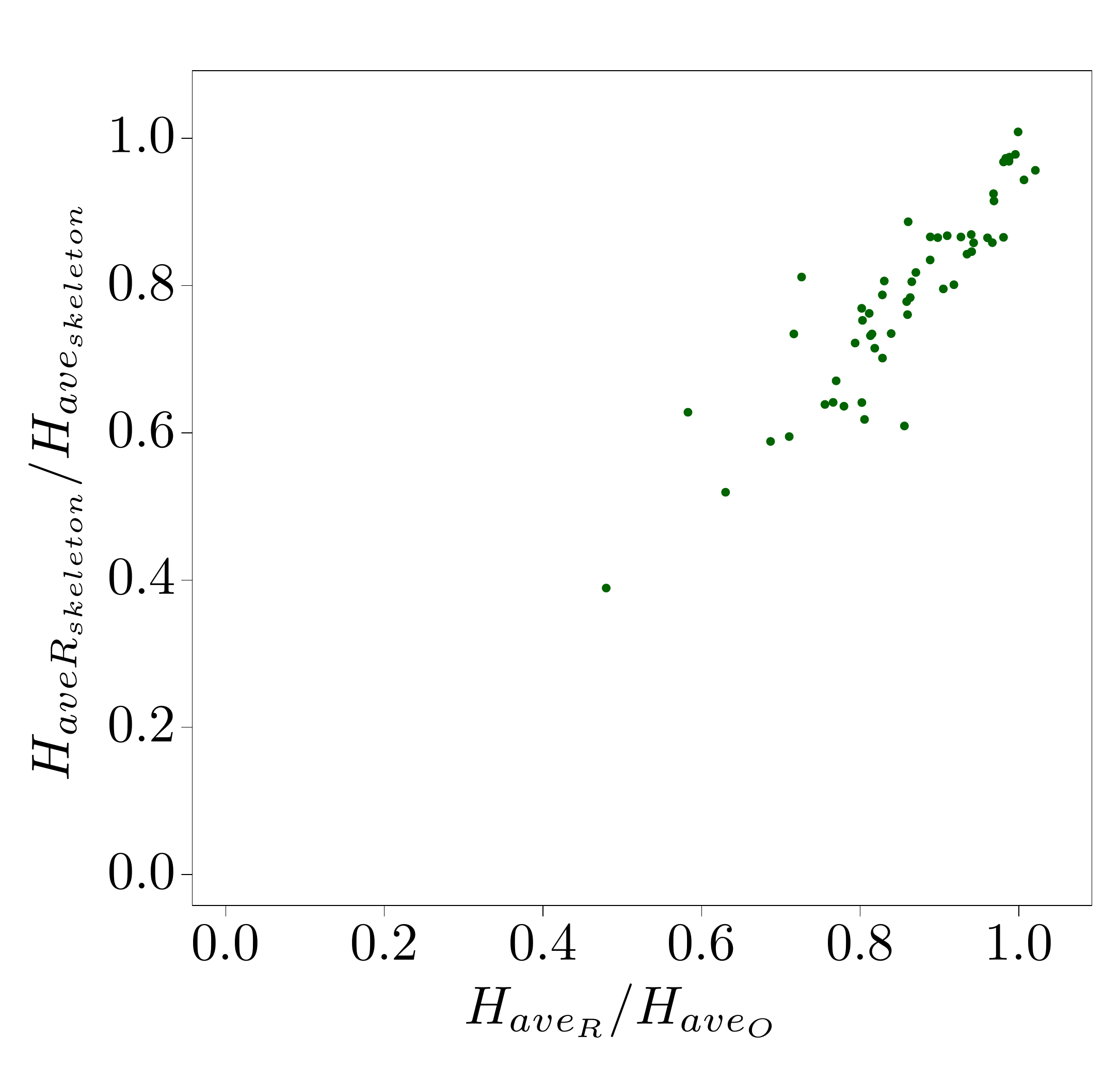}
\end{center}
\caption{\label{fig:eight}
The ratio of the average search information of the original network $H_{ave_O}$ and its random counterpart $H_{ave_R}$ against the ratio of the average search information of the skeleton network with the minimal search information obtained from the original network $H_{ave_{skeleton}}$ and its random counterpart $H_{ave_{Rskeleton}}$. All the random counterparts are strictly randomized following two rules: conserve the same degree distribution with the original network and ensure the connectivity of itself.
}
\end{figure}

Search information has been used to characterise how difficult is to navigate  a city. To take into account that cities have different sizes Rosvall~et~al. used the network average search information~\cite{Rosvall2005}
	\begin{equation}
	\hat\searchInfo=\frac{1}{N^2}\sum^N_{s=1}\sum^N_{d=1}\searchInfo(s,d).
	\end{equation}
They noticed that modern cities like Manhattan are easier to navigate than older cities like Ume$\accentset{\circ}{\rm a}$, i.e. $\hat H({\rm Manhattan})<\hat H({\rm Ume}\accentset{\circ}{\rm a})$. 
To decide if a city is difficult  to navigate or not, Rosvall~et~al. compared the average search information of the city against its random counterpart, where the random counterpart has the same degree distribution of the original network but not the geometrical constraints. This comparison indicates how easy is to find a destination in a networks. Rosvall~et~al. found out that many cities are more difficult to navigate that their random counterpart, i.e. $\hat\searchInfo >\hat\searchInfo_R$.  We extend their results and consider not only cities but many other real networks and investigate if the difficulty of navigating a real network is also captured in the skeleton networks. In this case we evaluate the skeleton network and its random counterpart. Figure~\ref{fig:eight} shows that for almost all networks and their skeletons,  as the ratios are bound in the unit square. Hence the skeleton and its randomised version also captures that real networks are more difficult to navigate that their random counterparts.

The second application is to approximate the search information for large networks.
For large networks the evaluation of their search information is challenging as it requires the evaluation of all shortest-paths, including degeneracies, for all source-destination pairs which, in general, is an expensive computational process.  
As the search information scales with the size of the network we use the tree–contraction to search for a skeleton network with a small number of nodes and then approximate the search information from the scaling
\begin{equation}
\frac{H_{sk}}{H_{o}} \sim 0.988 \left( \frac{N_{sk}}{N_{o}} \right)^{2.35}
\end{equation}
or using
\begin{equation}
\label{eq:searchAppro}
H_{o} \sim 1.012 \left( \frac{N_{sk}}{N_{o}} \right)^{-2.35}H_{sk}.
\end{equation} 

\begin{figure}
\begin{center}
\includegraphics[width=12cm]{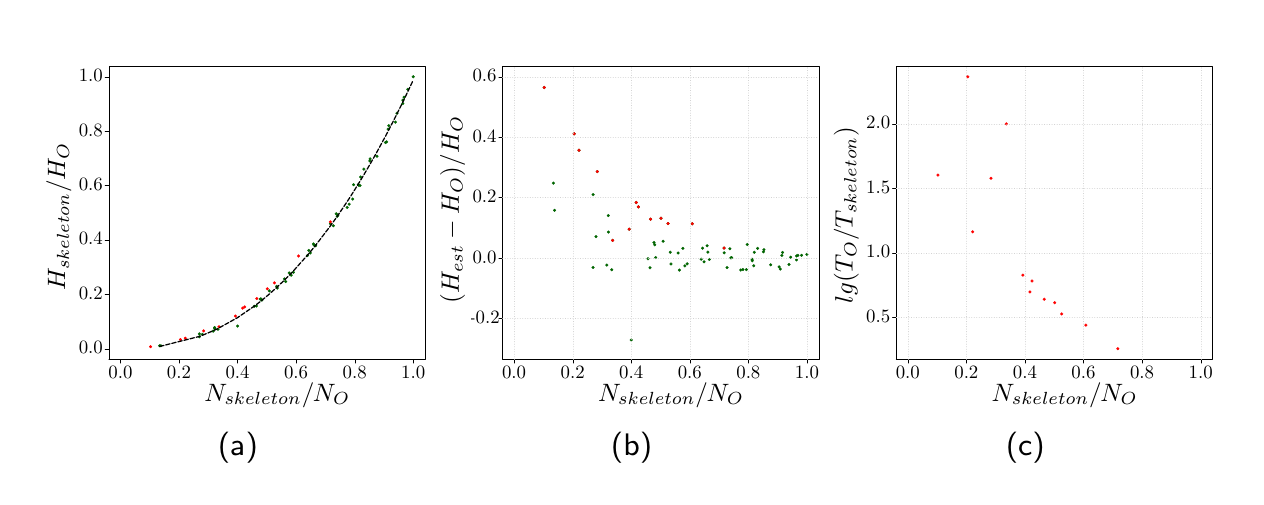}
\end{center}
\caption{\label{fig:seven}
(a)  Thirteen large networks (red dots) from different fields whcih fit the curve shown in Fig.~\ref{fig:five}(b). (b) Relative error of the approximation, and (c) shows the ratio of the computational time  of evaluating the search information $T_O$ relative to the time to evaluate the skeleton network $T_{skeleton}$ against $N_o/N_{skeleton}$. The green dots represent small networks which has less than 1000 nodes (shown in Supplementary information Tab.1) and red dots represent large networks which has more than 4000 nodes (shown in Supplementary information Tab.3)
}
\end{figure}

Figure~\ref{fig:seven}(a) shows the search information for  thirteen large networks which follow the scaling behaviour noticed in Fig.~\ref{fig:five}(b) indicating that Eq.~(\ref{eq:searchAppro}) can be use to approximate the search information of the original network. Fig.~\ref{fig:seven}(b) compares the relative error of the search information and its approximation as a function of the ratio $N_{sk}/N_o$. The approximation is good for large values of $N_{sk}/N_o$ but the error increases for small values of $N_{sk}/N_o$. The ratio  $N_{sk}/N_o$ is small if the number of nodes in the skeleton network is smaller that the number of nodes in the original network. This happens when the original network is more tree like, that is when the super-nodes contain large sized trees. In this case the approximation based on the scaling Eq.~(\ref{eq:searchAppro}) will be inaccurate. 
Fig.~\ref{fig:seven}(a) shows the ratio of the computational time between evaluating the search information for the original and the skeleton network against the the ratio  $N_{sk}/N_o$. Fig.~\ref{fig:seven}(c) shows that the computational time can be reduced by up to two orders of magnitude. As a rule of thumb, our method gives a reasonable approximation if $N_{sk}/N_o\ge 0.3$

\section*{Authors Contributions Statement}
HY, RGC and RJM carried all the research presented here, wrote the main manuscript text and prepared the figures. 


\begin{thebibliography}{10}

\bibitem{Fornito2016}
Chapter 7 - paths, diffusion, and navigation.
\newblock In Alex Fornito, Andrew Zalesky, and Edward~T. Bullmore, editors,
  {\em Fundamentals of Brain Network Analysis}, pages 207 -- 255. Academic
  Press, San Diego, 2016.

\bibitem{amico2019centralized}
Enrico Amico, Alex Arenas, and Joaqu{\'\i}n Go{\~n}i.
\newblock Centralized and distributed cognitive task processing in the human
  connectome.
\newblock {\em Network Neuroscience}, 3(2):455--474, 2019.

\bibitem{Avena-Koenigsberger2018}
Andrea Avena-Koenigsberger, Bratislav Misic, and Olaf Sporns.
\newblock Communication dynamics in complex brain networks.
\newblock {\em Nature Reviews Neuroscience}, 19(1):17--33, January 2018.

\bibitem{Barberillo2011}
J.~Barberillo and J.~Saldaña.
\newblock Navigation in large subway networks: An informational approach.
\newblock {\em Physica A: Statistical Mechanics and its Applications},
  390(2):374 -- 386, 2011.

\bibitem{1742-5468-2008-10-P10008}
V.~D. Blondel, Jean-Loup Guillaume, R.~Lambiotte, and E.~Lefebvre.
\newblock Fast unfolding of communities in large networks.
\newblock {\em Journal of Statistical Mechanics: Theory and Experiment},
  2008(10):P10008, 2008.

\bibitem{Boushaba2017}
Mustapha Boushaba, Abdelhakim Hafid, and Michel Gendreau.
\newblock Node stability-based routing in wireless mesh networks.
\newblock {\em Journal of Network and Computer Applications}, 93:1 -- 12, 2017.

\bibitem{JesperLinguistic}
Jesper Bruun.
\newblock {\em Networks in Physics Education Research}.
\newblock PhD thesis, Department Of Science Education Univeristy Of Copenhagen,
  2012.

\bibitem{Cajueiro2010}
Daniel~O. Cajueiro.
\newblock Optimal navigation for characterizing the role of the nodes in
  complex networks.
\newblock {\em Physica A: Statistical Mechanics and its Applications},
  389(9):1945 -- 1954, 2010.

\bibitem{PhysRevE.70.066111}
A.~Clauset, M.~E.~J. Newman, and C.~Moore.
\newblock Finding community structure in very large networks.
\newblock {\em Phys. Rev. E}, 70:066111, Dec 2004.

\bibitem{gallotti2016lost}
Riccardo Gallotti, Mason~A Porter, and Marc Barthelemy.
\newblock Lost in transportation: Information measures and cognitive limits in
  multilayer navigation.
\newblock {\em Science advances}, 2(2):e1500445, 2016.

\bibitem{Goni833}
Joaqu{\'\i}n Go{\~n}i, Martijn~P. van~den Heuvel, Andrea Avena-Koenigsberger,
  Nieves Velez~de Mendizabal, Richard~F. Betzel, Alessandra Griffa, Patric
  Hagmann, Bernat Corominas-Murtra, Jean-Philippe Thiran, and Olaf Sporns.
\newblock Resting-brain functional connectivity predicted by analytic measures
  of network communication.
\newblock {\em Proceedings of the National Academy of Sciences},
  111(2):833--838, 2014.

\bibitem{Liu:2009}
L.~Liu and R.~J. Mondragon.
\newblock {C}onservation of {A}lternative {P}aths {A}s a {M}ethod to {S}implify
  {L}arge {N}etworks.
\newblock In {\em Proceedings of the 1st Annual Workshop on Simplifying Complex
  Network for Practitioners}, page~1. ACM, 2009.

\bibitem{lloyd2018metro}
Peter~B Lloyd, Peter Rodgers, and Maxwell~J Roberts.
\newblock Metro map colour-coding: effect on usability in route tracing.
\newblock In {\em International Conference on Theory and Application of
  Diagrams}, pages 411--428. Springer, 2018.

\bibitem{Newman2004a}
M.~E.~J. Newman and M.~Girvan.
\newblock Finding and evaluating community structure in networks.
\newblock {\em Phys. Rev. E}, 69:026113, Feb 2004.

\bibitem{Perotti2012}
Juan Perotti and Orlando Billoni.
\newblock Smart random walkers: The cost of knowing the path.
\newblock {\em Physical Review E}, 86, 02 2012.

\bibitem{Pons2005}
Pascal Pons and Matthieu Latapy.
\newblock Computing communities in large networks using random walks (long
  version), 2005.

\bibitem{Rosvall1118}
M.~Rosvall and C.~T. Bergstrom.
\newblock Maps of random walks on complex networks reveal community structure.
\newblock {\em Proceedings of the National Academy of Sciences},
  105(4):1118--1123, 2008.

\bibitem{PhysRevE.72.046117}
M.~Rosvall, A.~Gr\"onlund, P.~Minnhagen, and K.~Sneppen.
\newblock Searchability of networks.
\newblock {\em Phys. Rev. E}, 72:046117, Oct 2005.

\bibitem{Rosvall2005}
M.~Rosvall, A.~Trusina, P.~Minnhagen, and K.~Sneppen.
\newblock {N}etworks and {C}ities: {A}n {I}nformation {P}erspective.
\newblock {\em Phys. Rev. Lett.}, 94:028701, Jan 2005.

\bibitem{Shahrezaye2019}
Morteza Shahrezaye, Orestis Papakyriakopoulos, Juan Carlos~Medina Serrano, and
  Simon Hegelich.
\newblock Measuring the ease of communication in bipartite social endorsement
  networks: A proxy to study the dynamics of political polarization.
\newblock In {\em Proceedings of the 10th International Conference on Social
  Media and Society}, SMSociety ’19, page 158–165, New York, NY, USA, 2019.
  Association for Computing Machinery.

\bibitem{Sneppen2005}
K~Sneppen, A~Trusina, and M~Rosvall.
\newblock Hide-and-seek on complex networks.
\newblock {\em Europhysics Letters ({EPL})}, 69(5):853--859, mar 2005.

\bibitem{stanley2018compressing}
Natalie Stanley, Roland Kwitt, Marc Niethammer, and Peter~J Mucha.
\newblock Compressing networks with super nodes.
\newblock {\em Scientific reports}, 8(1):10892, 2018.

\bibitem{Trusina2005}
A.~Trusina, M.~Rosvall, and K.~Sneppen.
\newblock Communication boundaries in networks.
\newblock {\em Phys. Rev. Lett.}, 94:238701, Jun 2005.

\bibitem{Zanin2008}
Massimiliano Zanin, Javier Buldu, Pedro Cano, and Stefano Boccaletti.
\newblock Disorder and decision cost in spatial networks.
\newblock {\em Chaos (Woodbury, N.Y.)}, 18:023103, 07 2008.

\bibitem{Zayani2012}
Mohamed-Haykel Zayani.
\newblock {\em {Link prediction in dynamic and human-centered mobile wireless
  networks}}.
\newblock Theses, {Institut National des T{\'e}l{\'e}communications}, September
  2012.

\end{thebibliography}
\end{document}


\flushbottom
{\Large
\begin{center}
Supplementary Information\\[2ex]
Simplification of networks by conserving path diversity \\and minimisation of the search information.\\[3ex]
{\normalsize H.~Yin, R.~G.~Clegg and R.~J.~Mondrag\'on}\\
\end{center}
}

All the networks datasets used here can be found in KONECT (the Koblenz Network Collection http://konect.uni-koblenz.de/) and ICON (The Colorado Index of Complex Networks https://icon.colorado.edu/\#!/).

\subsection*{Search information for some real networks and their simplification}
Table~\ref{tab:AllResults} shows the search information $\searchInfo_o$ for several real networks with $N_o$ nodes and its maximal and minimal search information when the networks are simplified.  The maxima search information $\searchInfo_{max}$ is for a skeleton with $N_{max}$ nodes and the minimal search information is $\searchInfo_{min}$ for a skeleton with $N_{min}$ nodes.

\begin{longtable}{| l || r | r || r | r || r | r |}

\hline
Network       & $N_o$ & $\searchInfo_o$  & $N_{max}$ & $\searchInfo_{max}$ & $N_{min}$ & $\searchInfo_{min}$ \\ \hline \hline
Bison         & 26    & 2904    & 26         & 2904       & 26        & 2904       \\ \hline
Cattle        & 28    & 3415    & 27         & 3078       & 27        & 3078       \\ \hline
Hens          & 32    & 4915    & 32         & 4915       & 32        & 4915       \\ \hline
Karate        & 34    & 6061    & 31         & 5078       & 29        & 4233       \\ \hline
USA           & 49    & 18093   & 46         & 15075      & 46        & 15075      \\ \hline
Dolphins      & 62    & 26082   & 39         & 8761       & 35        & 6469       \\ \hline
Polbooks      & 105   & 89776   & 93         & 66496      & 92        & 63529      \\ \hline
D.Copperfiled & 112   & 86869   & 66         & 26263      & 58        & 19369      \\ \hline
Football      & 115   & 94428   & 83         & 44867      & 74        & 34081      \\ \hline
FloEcoWet     & 128   & 112356  & 125        & 106205     & 124       & 103914     \\ \hline
FloEcoDry     & 128   & 112210  & 125        & 106067     & 124       & 103778     \\ \hline
JazzMusicians & 198   & 349259  & 191        & 319189     & 191       & 319189     \\ \hline
C.elegant         & 279   & 685719  & 240        & 489999     & 233       & 456838     \\ \hline
TFL           & 369   & 1508840 & 58         & 120362     & 51        & 31382      \\ \hline
Network Scientist        & 379   & 2615736 & 343        & 1983709    & 343       & 1983707    \\ \hline
Airport           & 500   & 2930302 & 399        & 1777967    & 398       & 1768273    \\ \hline
Taro          & 22    & 1909    & 16         & 1003       & 13        & 542        \\ \hline
Residence     & 217   & 388220  & 184        & 264908     & 178       & 245237     \\ \hline
Swedish       & 15    & 774     & 6          & 93         & 6         & 93         \\ \hline
LesMis        & 77    & 46293   & 57         & 22726      & 57        & 22724      \\ \hline
JPR           & 92    & 56563   & 72         & 31687      & 66        & 26048      \\ \hline
Teacher       & 60    & 22117   & 26         & 4052       & 19        & 1769       \\ \hline
Trainbomb     & 64    & 30353   & 50         & 16750      & 50        & 16139      \\ \hline
Friendship    & 22    & 2153    & 20         & 1638       & 20        & 1638       \\ \hline
Ecoli         & 329   & 1230895 & 102        & 142646     & 93        & 87239      \\ \hline
Gamethrones   & 107   & 94989   & 87         & 57223      & 87        & 57174      \\ \hline
New Spain     & 224   & 519626  & 31         & 112130     & 30        & 13937      \\ \hline
London Gang   & 54    & 17333   & 53         & 16533      & 53        & 16533      \\ \hline
Massachusetts & 74    & 42881   & 52         & 19310      & 48        & 15126      \\ \hline
Siren         & 44    & 12429   & 29         & 5646       & 29        & 5646       \\ \hline
911terrorist  & 62    & 28781   & 48         & 15525      & 48        & 15036      \\ \hline
Mexican       & 35    & 6067    & 32         & 4960       & 32        & 4898       \\ \hline
Blumenaudrug  & 75    & 35567   & 43         & 10141      & 36        & 6698       \\ \hline
Strike        & 24    & 2804    & 16         & 1149       & 16        & 1067       \\ \hline
Grassland     & 75    & 42988   & 41         & 12202      & 40        & 10095      \\ \hline
Rhodesbomb    & 22    & 2051    & 18         & 1241       & 18        & 1241       \\ \hline
Korea         & 33    & 6456    & 24         & 2927       & 24        & 2927       \\ \hline
Literary      & 35    & 5588    & 22         & 2035       & 16        & 897        \\ \hline
Heroin        & 38    & 7453    & 29         & 4184       & 28        & 3712       \\ \hline
Gang          & 29    & 4021    & 23         & 2217       & 23        & 2217       \\ \hline
Saltmarsh     & 128   & 115299  & 116        & 93734      & 109       & 79569      \\ \hline
SEP1999       & 56    & 22136   & 26         & 3985       & 26        & 3586       \\ \hline
Attiro        & 59    & 21392   & 41         & 9847       & 34        & 6014       \\ \hline
Yeast         & 662   & 5455250 & 233        & 558288     & 215       & 463329     \\ \hline
Sanjuansur    & 75    & 38830   & 49         & 14908      & 42        & 10105      \\ \hline
KBN           & 519   & 2818170 & 168        & 246851     & 138       & 159801     \\ \hline
Anaheim       & 416   & 2290732 & 147        & 235141     & 131       & 170660     \\ \hline
Russiatrade   & 39    & 7471    & 18         & 1761       & 13        & 673        \\ \hline
Mali          & 36    & 7838    & 21         & 2138       & 21        & 2138       \\ \hline
Cherowitzo    & 60    & 21856   & 33         & 5838       & 29        & 3979       \\ \hline
Italiangang   & 65    & 32393   & 43         & 12405      & 43        & 12380      \\ \hline
Prison        & 67    & 28083   & 38         & 8267       & 32        & 5217       \\ \hline
Basketball    & 56    & 18978   & 36         & 6865       & 30        & 4276       \\ \hline
Copilot       & 48    & 12988   & 41         & 9165       & 39        & 7856       \\ \hline
Stmarks       & 54    & 16426   & 51         & 14298      & 51        & 14225      \\ \hline
Drugnet       & 193   & 403421  & 55         & 34545      & 52        & 20839      \\ \hline
Sawmill       & 36    & 7056    & 26         & 3334       & 23        & 2430       \\ \hline
Maspalomas    & 24    & 2413    & 22         & 2029       & 22        & 1978       \\ \hline
Parisfootball & 35    & 5911    & 26         & 2980       & 26        & 2904       \\ \hline

\caption{ }
\label{tab:AllResults}
\end{longtable}

%
\newpage

\subsection*{The path diversity of different network simplification methods}
Table~\ref{tab1:cyclomaticComparison} shows the evaluation of the cyclomatic number under different clustering algorithms. The cyclomatic number is $C=L-N+P$ where $L$ is the number of links, $N$ the number of nodes and $P$ the number of connected components. 
\begin{longtable}{|l || c|c|c|c|c|c|c|}
\hline
Network           & $C_O$                                                     & $C_{TC}$                                                  & $C_{Lovian}$                                                    & $C_{FG}$                                               & $C_{IM}$                                                   & $C_{WT}$                                                  & $C_{BC}$ \\ \hline \hline
Karate            & \begin{tabular}[c]{@{}c@{}}45\\ (34,78)\end{tabular}      & \begin{tabular}[c]{@{}c@{}}45\\ (29,73)\end{tabular}      & \begin{tabular}[c]{@{}c@{}}27\\ (34,57)\end{tabular}      & \begin{tabular}[c]{@{}c@{}}28\\ (34,59)\end{tabular}      & \begin{tabular}[c]{@{}c@{}}33\\ (34,64)\end{tabular}      & \begin{tabular}[c]{@{}c@{}}17\\ (34,46)\end{tabular}      & \begin{tabular}[c]{@{}c@{}}25\\ (34,54)\end{tabular}      \\ \hline
Dolphins          & \begin{tabular}[c]{@{}c@{}}98\\ (62,159)\end{tabular}     & \begin{tabular}[c]{@{}c@{}}98\\ (38,135)\end{tabular}     & \begin{tabular}[c]{@{}c@{}}63\\ (62,120)\end{tabular}     & \begin{tabular}[c]{@{}c@{}}73\\ (62,131)\end{tabular}     & \begin{tabular}[c]{@{}c@{}}66\\ (62,116)\end{tabular}     & \begin{tabular}[c]{@{}c@{}}73\\ (62,131)\end{tabular}     & \begin{tabular}[c]{@{}c@{}}70\\ (62,127)\end{tabular}     \\ \hline
Polbooks          & \begin{tabular}[c]{@{}c@{}}337\\ (105,441)\end{tabular}   & \begin{tabular}[c]{@{}c@{}}337\\ (92,428)\end{tabular}    & \begin{tabular}[c]{@{}c@{}}299\\ (105,400)\end{tabular}   & \begin{tabular}[c]{@{}c@{}}302\\ (105,403)\end{tabular}   & \begin{tabular}[c]{@{}c@{}}278\\ (105,377)\end{tabular}   & \begin{tabular}[c]{@{}c@{}}302\\ (105,403)\end{tabular}   & \begin{tabular}[c]{@{}c@{}}299\\ (105,399)\end{tabular}   \\ \hline
D.Copperfield     & \begin{tabular}[c]{@{}c@{}}314\\ (112,425)\end{tabular}   & \begin{tabular}[c]{@{}c@{}}314\\ (62,375)\end{tabular}    & \begin{tabular}[c]{@{}c@{}}93\\ (112,198)\end{tabular}    & \begin{tabular}[c]{@{}c@{}}100\\ (112,205)\end{tabular}   & \begin{tabular}[c]{@{}c@{}}311\\ (112,421)\end{tabular}   & \begin{tabular}[c]{@{}c@{}}124\\ (112,211)\end{tabular}   & \begin{tabular}[c]{@{}c@{}}75\\ (112,118)\end{tabular}    \\ \hline
Football          & \begin{tabular}[c]{@{}c@{}}499\\ (115,613)\end{tabular}   & \begin{tabular}[c]{@{}c@{}}499\\ (78,576)\end{tabular}    & \begin{tabular}[c]{@{}c@{}}338\\ (115,444)\end{tabular}   & \begin{tabular}[c]{@{}c@{}}339\\ (115,448)\end{tabular}   & \begin{tabular}[c]{@{}c@{}}320\\ (115,423)\end{tabular}   & \begin{tabular}[c]{@{}c@{}}327\\ (115,432)\end{tabular}   & \begin{tabular}[c]{@{}c@{}}330\\ (115,435)\end{tabular}   \\ \hline
JazzMusicians     & \begin{tabular}[c]{@{}c@{}}2545\\ (198,2742)\end{tabular} & \begin{tabular}[c]{@{}c@{}}2545\\ (191,2735)\end{tabular} & \begin{tabular}[c]{@{}c@{}}1812\\ (198,2006)\end{tabular} & \begin{tabular}[c]{@{}c@{}}1943\\ (198,2137)\end{tabular} & \begin{tabular}[c]{@{}c@{}}2357\\ (198,2548)\end{tabular} & \begin{tabular}[c]{@{}c@{}}1967\\ (198,2154)\end{tabular} & \begin{tabular}[c]{@{}c@{}}1786\\ (198,1945)\end{tabular} \\ \hline
C.elegans         & \begin{tabular}[c]{@{}c@{}}2009\\ (279,2287)\end{tabular} & \begin{tabular}[c]{@{}c@{}}2009\\ (239,2247)\end{tabular} & \begin{tabular}[c]{@{}c@{}}1223\\ (279,1498)\end{tabular} & \begin{tabular}[c]{@{}c@{}}1488\\ (279,1763)\end{tabular} & \begin{tabular}[c]{@{}c@{}}1381\\ (279,1662)\end{tabular} & \begin{tabular}[c]{@{}c@{}}1422\\ (279,1698)\end{tabular} & \begin{tabular}[c]{@{}c@{}}1518\\ (279,1789)\end{tabular} \\ \hline
TfL               & \begin{tabular}[c]{@{}c@{}}62\\ (369,430)\end{tabular}    & \begin{tabular}[c]{@{}c@{}}62\\ (54,115)\end{tabular}     & \begin{tabular}[c]{@{}c@{}}34\\ (369,384)\end{tabular}    & \begin{tabular}[c]{@{}c@{}}37\\ (369,387)\end{tabular}    & \begin{tabular}[c]{@{}c@{}}26\\ (369,339)\end{tabular}    & \begin{tabular}[c]{@{}c@{}}36\\ (369,360)\end{tabular}    & \begin{tabular}[c]{@{}c@{}}30\\ (369,379)\end{tabular}    \\ \hline
Network Scientist & \begin{tabular}[c]{@{}c@{}}536\\ (379,914)\end{tabular}   & \begin{tabular}[c]{@{}c@{}}536\\ (343,878)\end{tabular}   & \begin{tabular}[c]{@{}c@{}}488\\ (379,849)\end{tabular}   & \begin{tabular}[c]{@{}c@{}}495\\ (379,855)\end{tabular}   & \begin{tabular}[c]{@{}c@{}}432\\ (379,770)\end{tabular}   & \begin{tabular}[c]{@{}c@{}}493\\ (379,846)\end{tabular}   & \begin{tabular}[c]{@{}c@{}}489\\ (379,850)\end{tabular}   \\ \hline
Airport           & \begin{tabular}[c]{@{}c@{}}2481\\ (500,2980)\end{tabular} & \begin{tabular}[c]{@{}c@{}}2481\\ (399,2879)\end{tabular} & \begin{tabular}[c]{@{}c@{}}1603\\ (500,2090)\end{tabular} & \begin{tabular}[c]{@{}c@{}}1573\\ (500,2062)\end{tabular} & \begin{tabular}[c]{@{}c@{}}2247\\ (500,2716)\end{tabular} & \begin{tabular}[c]{@{}c@{}}1905\\ (500,2364)\end{tabular} & \begin{tabular}[c]{@{}c@{}}2400\\ (500,2792)\end{tabular} \\ \hline

\caption{\label{tab1:cyclomaticComparison}
The path diversity of simplified networks obtained from tree-contraction method and 5 different graph partitioning methods. The upper row of $C_O$ is the cyclomatic number of the original networks. $C_{TC}$, $C_{Lovian}$, $C_{FG}$, $C_{IM}$, $C_{WT}$ and $C_{BC}$ are the cyclomatic number of the simplified networks obtained from the different simplification methods. The numbers in brackets are the number of nodes and edges between clusters.}
\end{longtable}  

\subsection*{Search information for some large networks and their simplifications}
Table~\ref{tab:AllResultsTwo} shows the search information $\searchInfo_o$ for several large networks with $N_o$ nodes and its maximal and minimal search information when the networks are simplified. $\searchInfo_{skeletonmax}$ is the maximal search information of the skeleton with $N_{skeletonmax}$ nodes and $\searchInfo_{skeletonmin}$ is the minimal search information of the skeleton with $N_{skeletonmin}$ nodes. 

\begin{longtable}{| l || c | c || c | c || c | c |}

\hline
Network       & $N_o$ & $\searchInfo_o$  & $N_{skeletonmax}$ & $\searchInfo_{skeletonmax}$ & $N_{skeletonmin}$ & $\searchInfo_{skeletonmin}$ \\ \hline \hline
Internet     & 11175 & 1898544710  & 5224   & 405367311  & 5192   & 399437235  \\ \hline
AS           & 10901 & 2079970369  & 5744   & 506240512  & 5720   & 501864031  \\ \hline
BGP          & 17447 & 5600982163  & 8769   & 1239334130 & 8728   & 1226782718 \\ \hline
Caida        & 26476 & 13635729216 & 11264  & 2103763543 & 11211  & 2082011965 \\ \hline
Chess        & 7116  & 792010798   & 2989   & 119923171  & 2951   & 116773785  \\ \hline
Chicago      & 12980 & 7912207191  & 4395   & 662173757  & 4369   & 605898315  \\ \hline
Google       & 23614 & 11238819758 & 6721   & 741892468  & 6717   & 729417488  \\ \hline
HostPathogen & 10862 & 2067554579  & 2253   & 70718006   & 2205   & 67440512   \\ \hline
Interactome  & 5583  & 473240887   & 1248   & 18630363   & 1224   & 17819455   \\ \hline
PGP          & 10681 & 2693004566  & 4205   & 325075319  & 4191   & 321274251  \\ \hline
SisterCity   & 10321 & 1872213563  & 1081   & 14204962   & 1041   & 13097524   \\ \hline
Skitter      & 9201  & 1333980676  & 5603   & 456250412  & 5587   & 453370967  \\ \hline
\caption{ Search Information}
\label{tab:AllResultsTwo}
\end{longtable}